\documentclass[12pt]{article}
\usepackage{doublespace}
\usepackage{graphicx}
\begin{document}

\centerline{\Large\bf The nature of gravitational singularities}
\bigskip
\bigskip
\centerline{{\bf David Garfinkle}\footnote{email address:
david@physics.uoguelph.ca}}
\bigskip
\bigskip
\centerline{Department of Physics, University of Guelph, Guelph, Ontario,
Canada N1G 2W1}
\centerline{ and Perimeter Institute for Theoretical Physics, 
35 King Street North, Waterloo Ontario, Canada N2J 2W9}
\bigskip

\begin{abstract}

The nature of gravitational singularities, long mysterious, has now
become clear through a combination of mathematical and numerical 
analysis.  As the singularity is approached, the time derivative terms
in the field equations dominate, and the singularity behaves locally
like a homogeneous oscillatory spacetime. 

\end{abstract}
\newpage

A longstanding problem in general relativity has been to determine the 
nature of the singularities that form in gravitational collapse.  
Powerful theorems due to Hawking, Penrose and others\cite{HandE} show
that singularities form under very general circumstances.  However, these
theorems give almost no information about the nature of singularities,
saying only that the worldline of some observer or light ray fails to be
complete.   There is also a longstanding conjecture due to Belinskii, 
Khalatnikov and Lifschitz (BKL)\cite{bkl} on the general nature of 
singularities.  The BKL conjecture is that as a general singularity
is approached, the dynamics becomes local and oscillatory.  The
analysis of reference\cite{bkl} was heuristic, so what was needed were 
two things: (1) a more precise way of stating the BKL conjecture and (2)
a way of checking whether it is true.     

It was realized by Berger and Moncrief\cite{bevandvince}
that numerical simulations provide
a way of checking the BKL conjecture: simulate the evolution of the
spacetime as the singularity is approached and see whether the behavior is
as conjectured in reference\cite{bkl}.  What resulted from this insight
was a program of research\cite{bevreview} that simulated the approach to
the singularity in spacetimes with symmetry.  The imposition of 
symmetry made the equations simpler and allowed the simulations (which 
were constrained by limits of computer memory) to be done with high 
spatial resolution.  Both the simulations and the analysis of the 
results were done by casting the Einstein field equations as a 
Hamiltonian system.   
The results supported the BKL conjecture: in all
cases the dynamics became local, {\it i.e.} spatial derivatives
in the field equations became negligible compared to time derivatives.     
In some cases the dynamics was oscillatory and in some cases not.  However,
one could plausibly argue that the cases where the dynamics was not 
oscillatory were not sufficiently general and that the general case would
be expected to be oscillatory.   

One limitation of this research program was the imposition of symmetry.
As long as spacetimes with symmetry were treated, one could never be
sure that the results reflected the behavior of the general spacetime
without symmetry.  Another difficulty came from the use of Hamiltonian
variables.  These were sufficiently different from the variables used
in reference\cite{bkl} that it was often difficult to compare the results
of the simulations to the expected BKL behavior.

These difficulties came to be resolved  
with the use of scale invariant variables in the work of Uggla {\it et al}
\cite{jw}.  Here the key insight comes from the scale invariance of
the vacuum Einstein equations, that is the property that a solution of 
these equations remains a solution if the overall length scale is changed.
In the homogeneous, isotropic spacetimes of big bang cosmology a scale 
(or more precisely a scale at a given time) is
given by the value of the Hubble constant.  This is the rate of the 
expansion of space.  The notion of the Hubble constant depends specifically
on the homogeneity and isotropy of space; however it can be generalized
to the case with no symmetry.  In cosmology, the Hubble constant is 
one third of
the divergence of the normal to the surfaces of homogeneity, which form
the cosmological surfaces of constant time.  In a general spacetime, given
a choice of time slicing one can define the Hubble parameter $H$ at a 
given spacetime point to be simply one third of
the divergence of the normal to the
constant time surface.  In physical terms, one can think of space as 
having three different rates of expansion (or contraction) in each of
three orthogonal directions.  The Hubble parameter is then defined to be
the average rate of expansion.  Given this scale, one can then divide all
other variables by (appropriate powers of) $H$ to make them scale invariant.
One important variable is the shear $\sigma _{\alpha \beta}$ whose 
eigenvalues give the differences in the rates of expansion of the three
orthogonal directions.  In the equations, one uses the related
scale invariant variable 
${\Sigma _{\alpha \beta}}\equiv {\sigma _{\alpha \beta}}/H$. 
Another important variable is $n_{\alpha \beta}$
which measures the failure of derivatives along orthogonal spatial 
directions to commute and is therefore related to the curvature of 
space.  The related scale invariant variable is 
${N_{\alpha \beta}}\equiv {n_{\alpha \beta}}/H$.          
As the singularity is approached, $H$ diverges.  However, the scale
invariant quantities remain finite.

This set of variables also gives rise to a natural prescription for 
decomposing spacetime into space and time: pick an initial time slice
and an orthonormal spatial frame on this slice.  Choose a time 
orientation so that the singularity is to the past and choose time
evolution towards the singularity to be motion by  
the amount $H^{-1}$
along the direction normal to the slice.  Finally Fermi-Walker 
transport the spatial frame along the time evolution.  This gives
rise to a time coordinate that goes to minus infinity as the 
singularity is approached.  Behavior near the singularity then 
becomes the $t\to -\infty$ behavior of solutions of the scale invariant
system.   This prescription then gives a rigorous way of stating the
BKL conjecture: as $t\to - \infty$ the spatial derivatives of the
scale invariant variables become negligible and the behavior at each
spatial point becomes that of an oscillatory homogeneous spacetime.

What remained was to perform numerical simulations of the system of
reference \cite{jw} for the general case of no symmetry,
to see whether the conjecture was correct.  
Such simulations were performed by the author.\cite{dg}
The results support the BKL conjecture.  As the singularity is 
approached, spatial derivatives become negligible.  Each spatial
point then behaves like a homogeneous universe.  But what is the
behavior of a homogeneous universe?  And which type of homogeneous
universe corresponds to the general behavior of singularities?
Homogeneous universes can be classified by their
symmetry groups.  BKL conjectured that the general behavior of 
singularities is locally like that of a Mixmaster universe: a
homogeneous universe whose
symmetry group is $SU(2)$.   

\begin{figure}
\includegraphics[scale=1.0]{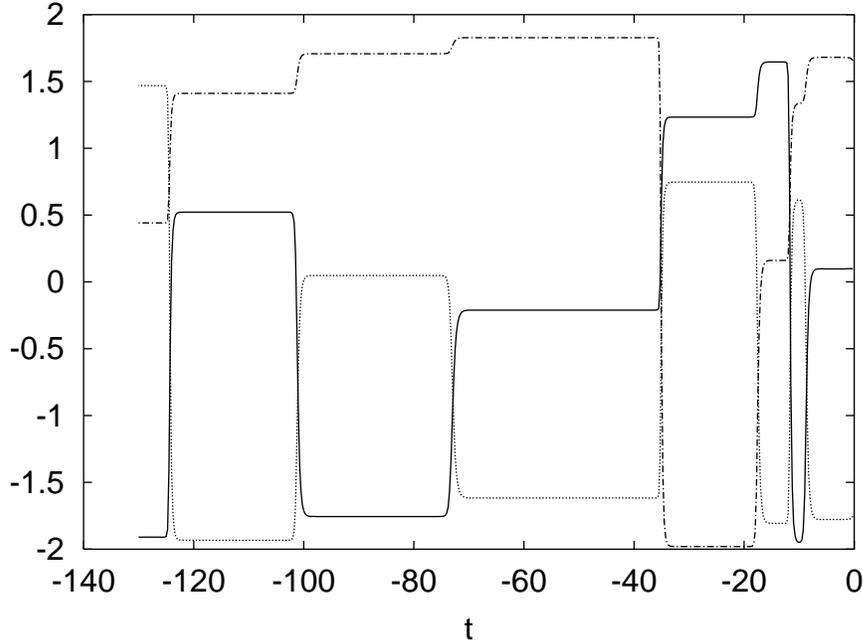}
\caption{\label{sigframe} components of $\Sigma_{\alpha \beta}$ 
{\it vs} time, in the 
asymptotic frame: $\Sigma _1$ (solid line), $\Sigma _2$ (dotted line)
and $\Sigma_3$ (dot-dashed line)}
\end{figure} 

\begin{figure}
\includegraphics[scale=1.0]{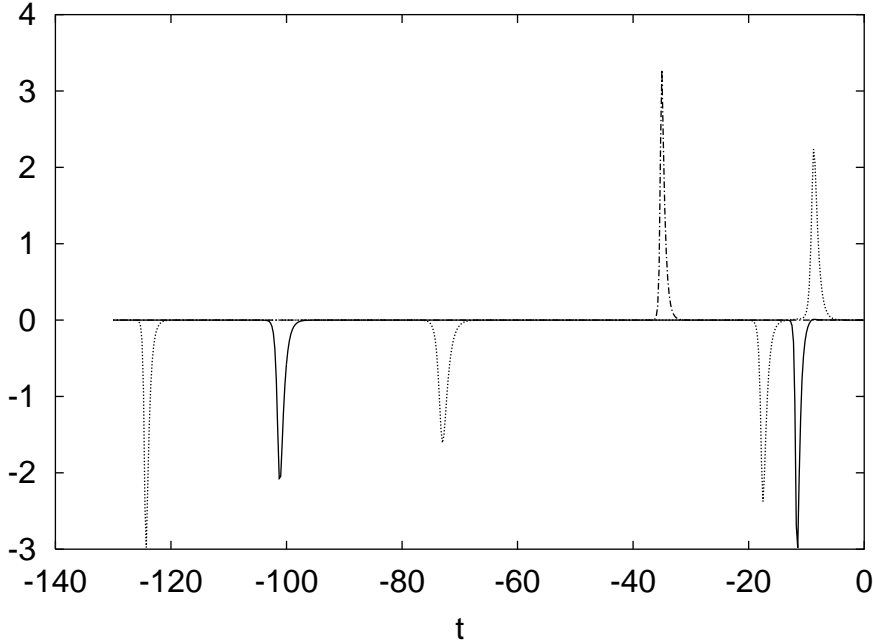}
\caption{\label{nframe} components of $N_{\alpha \beta}$ {\it vs} time, in the
asymptotic frame: $N_1$ (solid line), $N_2$ (dotted line) and $N_3$
(dot-dashed line)}
\end{figure}

To see whether this conjecture is true,
we must look at the dynamics of the scale invariant variables at
a single point.  Figures (\ref{sigframe}) and (\ref{nframe}) show
respectively the behavior of the variables $\Sigma _{\alpha \beta}$
and $N_{\alpha \beta}$ at a single spatial point in the numerical
simulation of reference\cite{dg} .  Here what is plotted are the
diagonal components of $\Sigma _{\alpha \beta}$ and $N_{\alpha \beta}$
in the ``asymptotic frame'' which is the frame of eigenvectors of
$\Sigma _{\alpha \beta}$ in the limit as the singularity is approached.
In the times between $0$ and $-20$ the spatial derivatives are not
negligible and the behavior is complicated.  However, for $t<-20$ the
spatial derivatives are negligible and the behavior is simple: it
consists of time intervals (called Kasner epochs) in which the components
of $\Sigma _{\alpha \beta}$ are constant while those of $N_{\alpha \beta}$
are negligible.  The Kasner epochs are punctuated by short ``bounces''
where during each bounce the components of $\Sigma _{\alpha \beta}$
change rapidly while one component of $N_{\alpha \beta}$ rapidly grows
and then decays.  This is exactly the behavior of a Mixmaster universe.

What determines how the components of $\Sigma _{\alpha \beta}$ change
from one Kasner epoch to the next?  To answer this, we must first characterize
the Kasner epochs.  The tensor $\Sigma _{\alpha \beta}$ is traceless, and 
during a Kasner epoch its square is equal to $6$.  Thus the three eigenvalues
of $\Sigma _{\alpha \beta}$ satisfy two relations and can therefore be
characterized by one parameter.  For each eigenvalue $\Sigma _i$
of $\Sigma _{\alpha \beta}$ introduce the number $p_i$ by 
${\Sigma _i}=3{p_i}-1$.  Then the properties of $\Sigma _{\alpha \beta}$
imply that during a Kasner epoch the sum of the $p_i$ and the sum of their
squares is $1$.  Now define $u$ to be the ratio of the largest $p_i$ to
the second largest one.  Then $u \ge 1$ and since there are two relations
satisfied by 3 $p_i$ it follows that the $p_i$ are completely characterized
by $u$.  The question of how $\Sigma _{\alpha \beta}$ changes from one
Kasner epoch to the next then becomes the question of how $u$ changes from
one epoch to the next.  For Mixmaster spacetimes the answer was found in
reference\cite{bkl}.  If $u\ge 2$ in one epoch, then it changes to 
$u-1$ in the next; while if $u\le 2$ in one epoch, then it changes to 
$1/(u-1)$ in the next.  This rule is called the $u$ map.   One can
compare the sequence of values of $u$ found in the simulations to the
rule of the $u$ map.  The result is that the general singularity satisfies
the $u$ map.  Thus the general singularity is local and oscillatory, with
the oscillations having the Mixmaster form.

Note that the part of the $u$ map of the form $u\to 1/(u-1)$ depends 
sensitively on initial conditions.  Thus the $u$ map and therefore
the general singularity are chaotic.  Nearby spatial points begin
with nearby values of $u$; but after a certain number of bounces they
have very different values of $u$.  

Finally note that the treatment of this essay has used classical 
general relativity; but the approach to the singularity necessarily
involves growth of curvature up to the Planck scale where classical
general relativity is no longer valid.  Though the $u$ map involves
an infinite sequence of bounces; the Planck scale will be reached
in a finite number of bounces.  The results of this paper should 
therefore be read as describing the approach to the Planck 
scale in gravitational
collapse.  After the Planck scale is reached, a description of 
gravitational collapse requires the calculations of a quantum theory
of gravity. 

I would like to thank Mark Miller, Beverly Berger, Woei-Chet Lim,
John Wainwright, Lars Andersson, Jim Isenberg and G. Comer Duncan
for helpful discussions.  This work was partially supported by
NSF grant PHY-0244683 to Oakland University.

\end{document}